\documentclass[letterpaper, 10 pt, conference]{ieeeconf}  

\IEEEoverridecommandlockouts                              
\usepackage{graphics} 
\usepackage{epsfig} 
\usepackage{amsmath} 
\usepackage{amssymb}  

\usepackage{url}
\usepackage[ruled, vlined, linesnumbered]{algorithm2e}
\usepackage{verbatim} 
\usepackage{soul, color}
\usepackage{lmodern}
\usepackage{fancyhdr}
\usepackage[utf8]{inputenc}
\usepackage{fourier} 
\usepackage{array}
\usepackage{makecell}

\SetNlSty{large}{}{:}

\makeatletter

\newcommand{\Rom}[1]{\expandafter\@slowromancap\romannumeral #1@}
\makeatother

\title{\LARGE \bf
Noise Reduction and Driving Event Extraction Method for Performance Improvement on Driving Noise-based Surface Anomaly Detection
}

\author{YeongHyeon Park$^{*}$\thanks{$^*$Corresponding author}, JoonSung Lee, Myung Jin Kim, Wonseok Park 
\\ SK Planet Co., Ltd. \\
{\tt\small\{yeonghyeon, js43.lee, myungjin, pwswonder\}@sk.com} \\
}

\begin{document}

\maketitle
\thispagestyle{plain}
\pagestyle{plain}

\begin{abstract}
Foreign substances on the road surface, such as rainwater or black ice, reduce the friction between the tire and the surface. The above situation will reduce the braking performance and make difficult to control the vehicle body posture. In that case, there is a possibility of property damage at least. In the worst case, personal damage will be occured. To avoid this problem, a road anomaly detection model is proposed based on vehicle driving noise. However, the prior proposal does not consider the extra noise, mixed with driving noise, and skipping calculations for moments without vehicle driving. In this paper, we propose a simple driving event extraction method and noise reduction method for improving computational efficiency and anomaly detection performance.

\end{abstract}


\section{Introduction}
\label{sec:intro}
The fundamental of the vehicle moving forward is based on the law of action and reaction. At the first, the power system of the vehicle rotates the wheel and the tire rotates with them. Then, the tire generates friction with the road surface. At this time, if the frictional force is higher than a specific level, the vehicle will move according to the direction of the wheel rotation. 

\begin{figure}
    \begin{center}
         \includegraphics[width=0.95\linewidth]{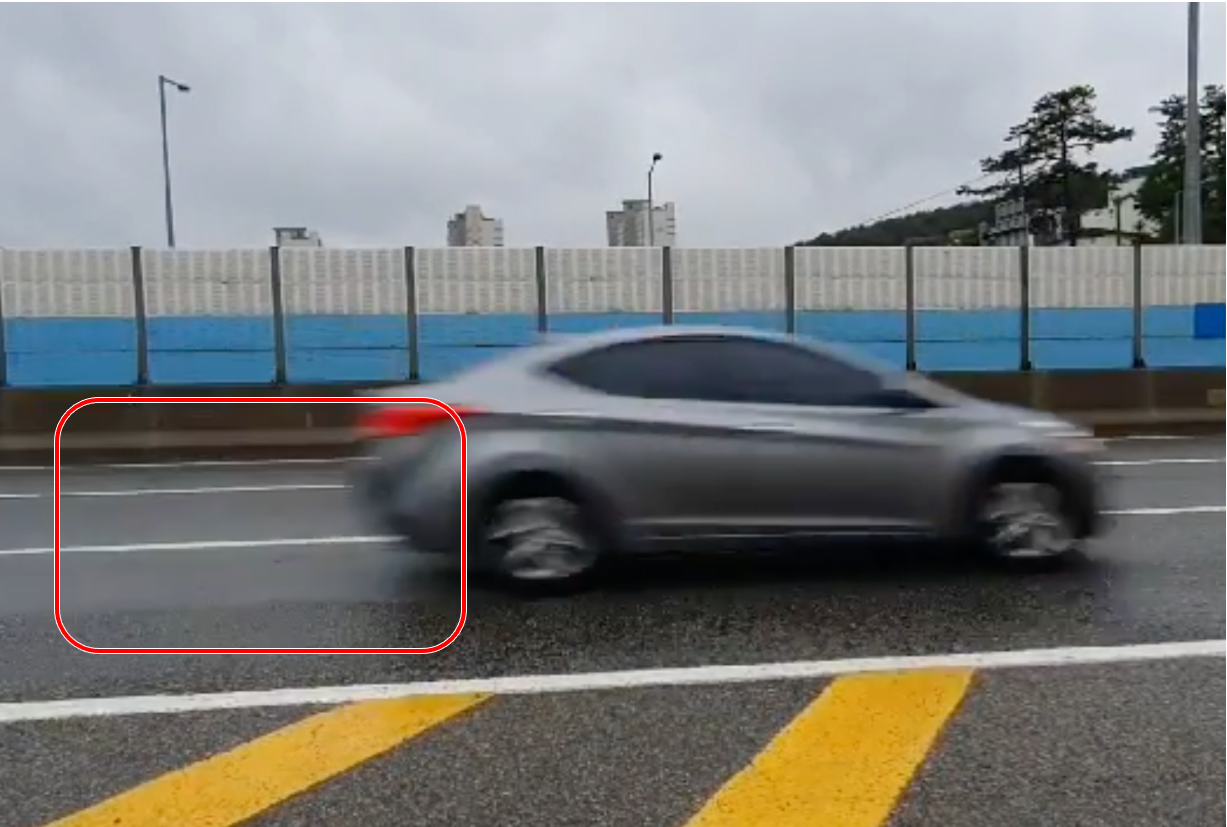}
	\end{center}
	\caption{Vehicle on wet roads. The road surface and the tire rub each other. The water fog, makred with red box, is generated by the water film crowded out from the road.}
	\label{fig:wetsurface}
\end{figure}

On the other hand, when foreign substances such as water or ice, the friction between the tires and the road surface will be reduced. Thus, in the above case driver cannot control the vehicle movement or braking immediately. The interference between the tires and the road surface in the case of the wet condition is shown in Fig.~\ref{fig:wetsurface}.

The coefficient of the friction between the tire and the road surface on a wet or icy road compared to a dry condition decreases from a minimum of 31\%-level to a maximum of 14\%-level \cite{hippi2010statistical}. Moreover, the probability of a serious accident is occured 1.3 times more than a dry condition, when the probability of an accident is investigated as 25\% on a dry road \cite{wallman2001friction}.

We have developed a model NCAE for driving noise-based anomaly detection to ease the above problem \cite{park2021non}. The NCAE shows an improvement in the anomaly detection performance compared to the prior models while increasing the computational efficiency.

However, when using the above model alone, performance changes according to weather in an outdoor environment. In particular, fickle noise such as wind noise disturbs the propagation of driving noise to the microphone in the air \cite{gutenberg1942propagation}. The baseline of the sound will fluctuate, and the wind noise may be misrecognized as a main characteristic of the acquired sound.

In this paper, to overcome the above limitations, we propose a noise reduction method for minimizing extra noise than driving noise. Also, the driving event extraction method is proposed to skip the calculation for moments without vehicles. For the verification of the proposed methods, we conduct experiments with two perspectives. One of them is the accuracy of driving event extraction and the other one is an improvement of anomaly detection performance.

\section{Proposed approach}
\label{sec:approach}
This section describes how to reduce extra noise than driving noise. In the process, an analysis of the characteristics of driving noise is also performed.

\begin{figure*}
    \begin{center}
         \includegraphics[width=0.95\linewidth]{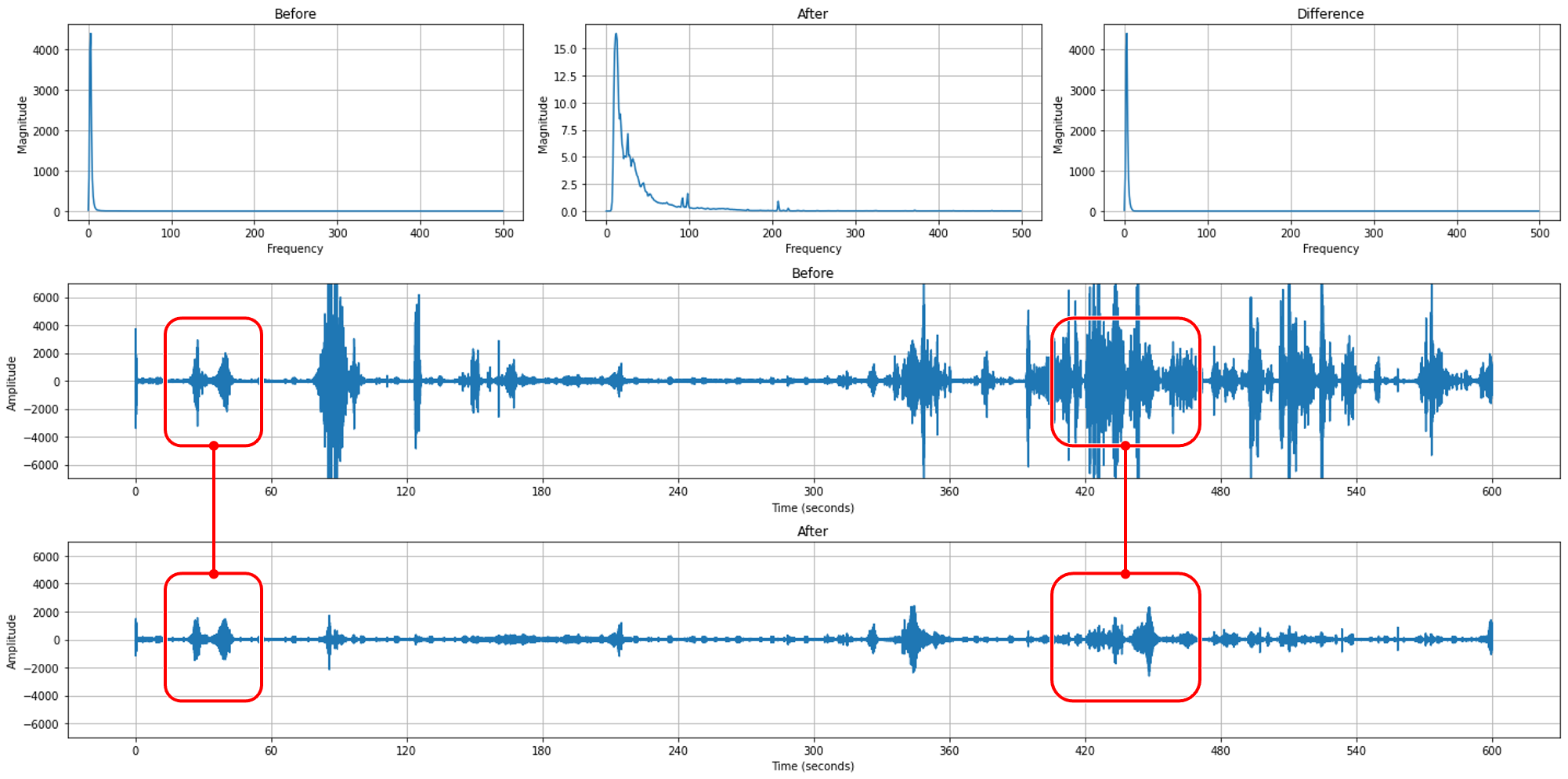}
	\end{center}
	\caption{The process for extracting driving events with performing noise reduction. In the first row, the time-averaged spectrum of the original sound data, after noise reduction, and the difference between the above two are sequentially shown. The second row and third row show the amplitude of the original sound data and data after noise reduction.}
	\label{fig:processing}
\end{figure*}

\section{Method for event extraction}
\label{subsec:methodextraction}
Determine the surface is normal or not by only vehicle driving moments is the purpose of extracting the driving event. To achieve the above, one of the importance is extracting driving events without missing them. The other importance is minimizing the situation in which noise, not a driving event, is considered and extracted as a driving event.

To extract driving events, we propose a concise and powerful method based on peak detection. First, smoothing is performed via the Hanning window on the sound data as shown in the second row of Fig.~\ref{fig:processing} on the time axis \cite{essenwanger1986elements}. Then, we consider the driving noise and extract, if the width of the peak exceeds the threshold and continues mote than 50ms on the time axis. The threshold is set by the lower 10\% of the sound data.

\section{Noise reduction}
\label{subsec:noisereduction}
Various noises are a hindrance factor for vehicle driving noise detection and extraction because our method is constructed based on a peak. In Fig.~\ref{fig:processing}, some of the actual driving noise is indicated by a red box.

As shown in the second row of Fig.~\ref{fig:processing}, when the amplitude of the original data is visualized, there is a section showing a prominent peak amplitude other than the section considered as some events. However, a high ratio of wind noise is mixed in that section. There is also a case it may be mistaken for driving noise in the signal view, but it actually wind noise.

The above characteristics occur in the lower 0.03\% and the upper 22\% region dominantly in the frequency domain. On the other hand, the driving noise is mostly located inside of the boundary. Therefore, we reduce the frequencies, which disrupts to determine the road surface condition as mentioned above bound, through the band-pass filter.

\pagebreak
\section{Experiments}
\label{sec:experiments}
We construct a dataset including tire and road friction noise for a vehicle driving on various road conditions. We use the above dataset for experiments to confim the effectiveness of proposed methods.

\subsection{Driving event extraction}
\label{subsec:eventextraction}
The dataset includes four types of sound recoreds dry, slush, snow, and wet. We have not collect icy road surface data due to weather and safety issues. The time length of the each record is 10 minutes equally.

In order to confirm the effectiveness of the noise reduction method, we apply the driving event extraction method for each record. The extraction result is summarized in Table~\ref{tab:event}. 

Before applying noise reduction, the ratio of the driving events from the extracted events is 66\%. In contrast, the ratio of the driving events is increased to 98\% ratio when the noise reduction method is applied.

\begin{table}[ht]
    \centering
    \caption{Comparison of event extraction accuracy}
    \begin{tabular}{c|cc}
        \hline
            \textbf{} & \textbf{Original} & \textbf{Noise Reduction} \\       
        \hline
            Extracted Events & 186 & 126 \\ 
        \hline
            Driving events & 122 (66\%) & 124 (98\%) \\ 
            Other events & 64 (34\%)& 2 (2\%) \\ 
        \hline
    \end{tabular}
    \label{tab:event}
\end{table}

\subsection{Anomaly detection}
\label{subsec:anomaly}
In order to confirm the influence of the noise reduction method, we use the extracted 124 driving events shown in Table~\ref{tab:event}. Even if the data is defined as a driving event, extra noises including wind noise are mixed because noise reduction is not been performed yet.

\begin{figure}
    \begin{center}
         \includegraphics[width=0.95\linewidth]{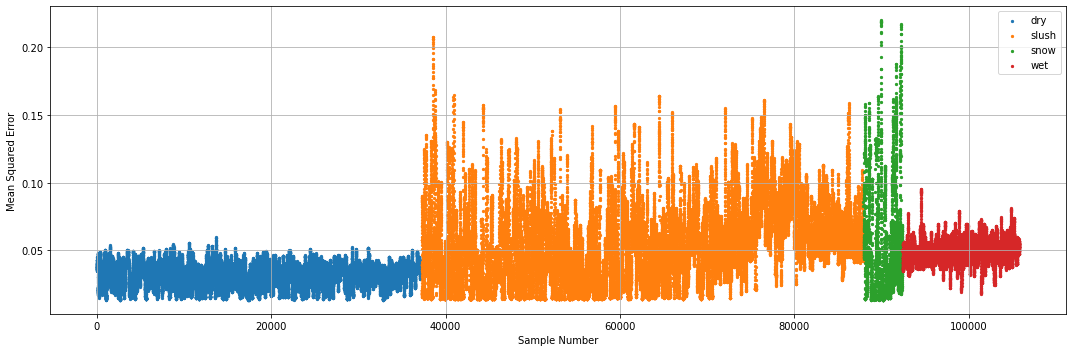} \\
         \includegraphics[width=0.95\linewidth]{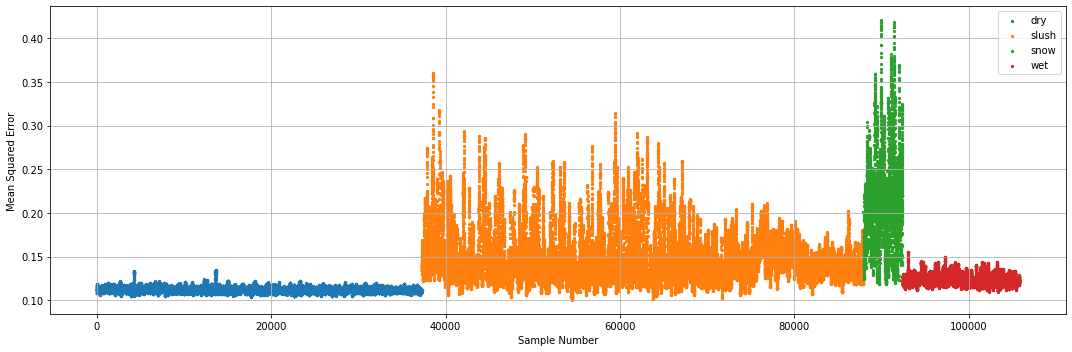} \\
	\end{center}
	\caption{The mean square error for the four cases of the road surface. Each type of sound is recorded with the same length but the number of the extracted driving event is different that depending on traffic load. The blue, orange, green, and red colors represent dry, slush, snow, and wet conditions respectively.}
	\label{fig:comparison2}
\end{figure}

We calculate the output via NCAE for two cases, original and noise reduction. Then, we measure and visualize the mean square error (MSE) between input and output of NCAE in Fig.~\ref{fig:comparison2}. When the noise reduction method is applied, the range of MSE for driving events corresponding to the normal is narrowed. Referring to the above situation reduces the risk of misjudging the normal state as abnormal.

We also compare the anomaly detection performance assuming summer and winter conditions, respectively. In summer, it is assumed that there are only two road surfaces, dry and wet. In winter, we use all four types dry, slush, snow, and wet as mentioned above.

We adopt the receiver operating characteristic curve (AUROC) as a performance indicator and the measured performance is summarized in Table~\ref{tab:anomalymulti}. Referring to Table~\ref{tab:anomalymulti}, the AUROC is improved by 12\% and 26\% for summer and winter conditions via applying the noise reduction method.

\begin{table}[ht]
    \centering
    \caption{Comparison of anomaly detection performance whether noise reduction is applied or not}
    \begin{tabular}{l|cc|c}
        \hline
            \textbf{} & \textbf{Original} & \textbf{Noise Reduction} & \textbf{Improvement}  \\       
        \hline
            Summer & 0.88351 & \textbf{0.98997} & 12\% \\
            Winter & 0.78143 & \textbf{0.98395} & 26\% \\
        \hline
    \end{tabular}
    \label{tab:anomalymulti}
\end{table}

\section{Conclusion}
\label{sec:conclusion}
In this paper, we propose an event extraction method that extracts driving events from a sound source. Also, a noise reduction method is proposed to improve the anomaly detection performance by suppressing the range of anomaly scores of the normal category. 

\pagebreak
The driving event extraction method helps to avoid unnecessary costs for calculating a period in which noise is not generated. In addition, we prove the noise reduction method is useful for suppressing noise other than driving sound. The accuracy of the driving event extraction is increased via noise reduction. Also, the anomaly detection performance for two season assumptions, summer and winter, is improved up to 26\% than the original sound data.

\section*{Acknowledgements}
We are grateful to all the members of SK Planet Co., Ltd., who have supported this research, providing equipment for the experiment.


\end{document}